\documentclass[pre,a4paper,superscriptaddress,twocolumn,showpacs,amsmath,amssymb,floatfix]{revtex4}

\usepackage{graphicx}
\usepackage{amssymb}
\usepackage{float}

\usepackage{color}

\input{epsf}

\begin{document}

\title{Dynamics and thermalization of Bose-Einstein condensate in Sinai oscillator trap}

\author{Leonardo Ermann}
\affiliation{
   Departamento de F\'{\i}sica, Gerencia de Investigaci\'on y Aplicaciones,
 Comisi\'on Nacional de Energ\'{\i}a At\'omica.
 Av.~del Libertador 8250, 1429 Buenos Aires, Argentina}
\affiliation{CONICET, Godoy Cruz 2290 (C1425FQB) CABA, Argentina}
\author{Eduardo Vergini}
\affiliation{
   Departamento de F\'{\i}sica, Gerencia de Investigaci\'on y Aplicaciones,
 Comisi\'on Nacional de Energ\'{\i}a At\'omica.
 Av.~del Libertador 8250, 1429 Buenos Aires, Argentina}
\author{Dima L. Shepelyansky}
\affiliation{\mbox{Laboratoire de Physique Th\'eorique, IRSAMC, 
Universit\'e de Toulouse, CNRS, UPS, 31062 Toulouse, France}}

\date{May 9, 2016}

\begin{abstract}
We study numerically the evolution of Bose-Einstein condensate 
in the Sinai oscillator trap described by the Gross-Pitaevskii
equation in two dimensions. 
In the absence of interactions this trap mimics the properties of Sinai
billiards where the classical dynamics is chaotic and 
the quantum evolution is described by 
generic properties of quantum chaos and random matrix theory.
We show that, above a certain border, the nonlinear interactions  
between atoms lead to the emergence of dynamical thermalization
which generates the statistical Bose-Einstein distribution
over eigenmodes of the system without interactions.
Below the thermalization border the evolution 
remains quasi-integrable.
Such a Sinai oscillator trap, formed
by the oscillator potential and a repulsive disk 
located in the vicinity of the center, had been already realized in
first experiments with the Bose-Einstein condensate 
formation by Ketterle group in 1995 and we argue that
it can form a convenient test bed for experimental
investigations of dynamical of 
thermalization. Possible links and implications
for Kolmogorov turbulence in absence of noise are also discussed.
\end{abstract}


\maketitle

\section{Introduction}

One of the first experimental realizations of Bose-Einstein condensate (BEC)
has been done with sodium atoms trapped in a novel trap 
that employed both magnetic and optical forces \cite{ketterle1995}. 
In this trap, the 
repulsive optical potential is created by tightly focusing 
an intense blue-detuned laser that generates a repulsive 
optical plug 
bunging a hole in a center of magnetic trap 
where nonadiabatic spin flips lead to a loss
of atoms. Further developments of BEC traps,
and remarkable progress of BEC experimentstheory
are reviewed in \cite{ketterle2002,ketterle2002rmp,pitaevskii,becbook}.

In spite of these achievements, the fundamental question about
interplay of dynamics, interactions and thermalization of BEC 
in a concrete trap configuration still waits its clarification.
In this work we address this question
in the frame of the Gross-Pitaevskii equation (GPE) \cite{pitaevskii,becbook}
for the two-dimensional (2D) version of the trap used in 
the experimental setup \cite{ketterle1995}.
Thus the trap potential is represented by a 2D harmonic potential
and a rigid disk which center is located in a 
vicinity of the  center of harmonic potential.
If the harmonic potential is replaced by rigid walls
forming a square or rectangle then the classical dynamics 
in such a Sinai billiard is proven to be completely chaotic
\cite{sinai1970}. The recent analysis of the trap with the 
walls formed by a harmonic potential shows that the dynamics
remains chaotic with a very small measure of 
integrable dynamics \cite{ermannepl}.
This trap was called the Sinai oscillator \cite{ermannepl}
due to a similarity with a Sinai billiard.
Since the realization of rigid walls is rather difficult
for experimental realization the case of Sinai oscillator trap
becomes much more attractive for 
combined theoretical and experimental investigations.
In fact a Sinai oscillator trap in three-dimensions
(3D) had been implemented in \cite{ketterle1995}.
Here we restrict our investigations to the 2D case
expecting that its main features will be preserved in 3D.

The quantum properties of Sinai oscillator
are characterized  within random matrix theory \cite{wigner}
by the Wigner-Dyson statistics of energy levels \cite{bohigas}.
The properties of the eigenstates are typical for
those of systems of quantum chaos and now are well understood 
(see e.g. \cite{haake,stockmann}).

Below we show that the Sinai oscillator trap
in 2D is also characterized by the properties of quantum chaos: 
the quantum eigenstates of Sinai oscillator 
are ergodic and the level spacing statistic is 
described by the random matrix theory,
in agreement with the Bohigas-Giannoni-Schmit conjecture \cite{bohigas,haake}.
However, still there is no thermalization in this system
since the eigenstates are preserved in
absence of interactions.
Thus, in this work we analyze the dynamics and thermalization conditions
for BEC in the Sinai oscillator
in the frame of the GPE equation. 
The GPE description is valid in the regime 
where the BEC temperature $T$ 
is below the critical temperature of 
Bose-Einstein condensation $T_c$ \cite{landau}
and when the validity of  GPE description
is well justified \cite{pitaevskii,becbook}.

Even though from the mathematical view point
the question about existence of solutions
of the GPE in such a trap, 
at moderate nonlinearity and large times,
remains an open problem (see e.g. \cite{kuksin,kuksin2015}). 
Indeed, the GPE can be rewritten in the basis of 
linear eigenstates (modes)
where the coupling between modes takes
place only due to the nonlinearity 
in GPE. In this representation each mode can be considered as an
independent oscillator degree of freedom 
and in case of thermalization, induced by nonlinearity,
one should expect energy equipartition
over all modes \cite{landau} 
leading to ultra-violet catastrophe 
and energy transfer to high energy modes.
In fact, the Planck constant and the Planck law 
had been introduced
for a black-body radiation to avoid such a divergence \cite{planck}.
However, the Planck distribution is valid for quantum
systems while in our case of the GPE Sinai
oscillator there is no second quantization. Thus due to
classical  
nonlinear interactions between modes one would expect
to have a classical ergodicity with 
equipartition of energy between modes.

Indeed, such an equipartition expectation was
at the origin of the studies of the Fermi-Pasta-Ulam (FPU) problem
\cite{fpu1,fpu2}. Nevertheless 
its absence is consequence of the 
proximity to the integrable Toda lattice 
(see e.g. \cite{fpu3,fpu4} and Refs. therein).  
Thus the FPU oscillator chain has certain specific features
which break system ergodicity in energy.
However, it is natural to expect that
in a generic case, when eigenstates of a linear system 
are ergodic and
dynamical chaos of classical trajectories takes place,
the energy equipartition over modes 
should appear above certain border of 
nonlinear interaction strength
between modes.

In spite of these expectations of 
energy equipartition over modes,
the recent studies of the GPE in Bunimovich stadium
showed that the nonlinearity
produces an effective dynamical thermalization
in a completely isolated system,
without any contact with external thermal bath,
with the probabilities over linear modes
described by the Bose-Einstein (BE) distribution  \cite{ermannepl}.
Thus the probabilities on high energy modes
drop rapidly and the ultra-violet catastrophe is absent.
An experimental realization of the Bunimovich
billiard with cold atoms is possible but is not so simple.
Due to this reason we consider here the GPE Sinai
oscillator trap which in fact has been already
built in \cite{ketterle1995,ketterle2002,ketterle2002rmp}
but without investigation of phenomenon of dynamical thermalization.
Our results show the presence of dynamical thermalization 
with BE distribution in this system
even if some aspects still should be clarified for
time evolution of very large time scales.

The model description, the quantum chaos features
of linear system are described in Section II. 
The thermalization equations and the formalism 
are presented in Section III.
The  obtained numerical
results for  the GPE evolution are presented and discussed 
in the Section IV. 
Numerical methods and behavior on large time scales are
discussed in Section V.
The discussion of the main results is presented in Section VI.


\section{Model description and quantum chaos properties}


The dynamics of the classical Sinai oscillator
 is described by the Hamiltonian:
\begin{equation}
 H=\frac{1}{2m}(p_x^2+p_y^2)+
\frac{m}{2}(\omega_x^2 x^2+\omega_y^2 y^2)+V_d(x,y) \;\; ,
\label{eq1}
\end{equation}
with the first two terms being 2D oscillator with 
frequencies $\omega_x, \omega_y$, while the last term describes
the potential wall of elastic disk of radius  $r_d$.
In our studies we fixed the mass $m=1$, 
frequencies $\omega_x=1$, $\omega_y=\sqrt{2}$
and disk radius $r_d=1$. The disk center is placed at
 $(x_d,y_d)=(-1/2,-1/2)$ so that the disk bungs a hole in the center
as it was the case in the experiments \cite{ketterle1995}.

\begin{figure}[t]
\begin{center}
\includegraphics[width=0.48\textwidth]{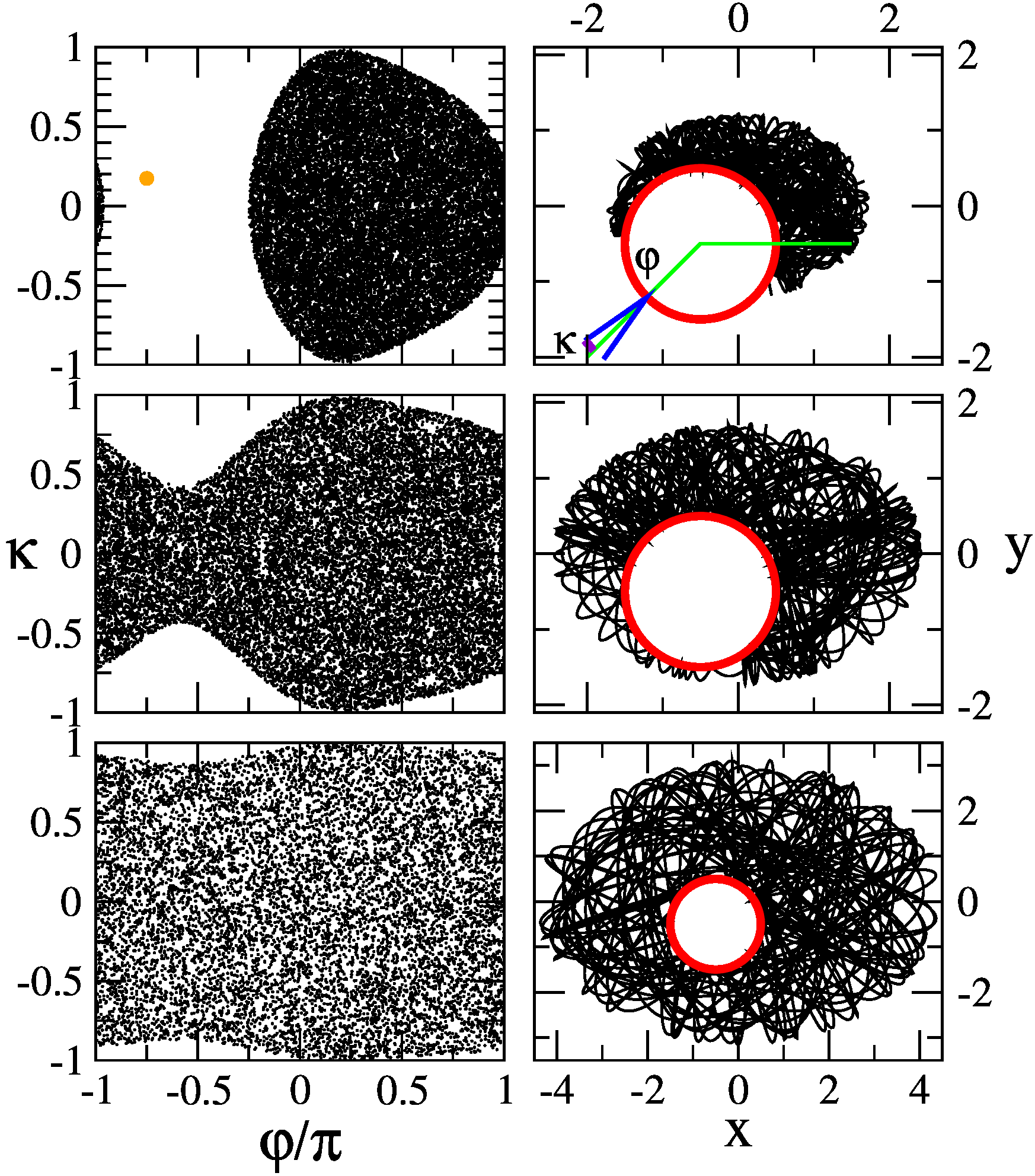}
\end{center}
\vglue -0.3cm
\caption{\label{fig1}
(Color online) Left panels show  
Poincar\'e section given by canonical variables $\varphi$ and $\kappa$,
taken at the disk bounce;
points show trajectories evolving up to time
$t=10000$ with $5$ random  initial conditions 
at initial energies $E=1.5; 3; 10$ in top, 
center and bottom panels respectively.
Right panels show  dynamics of one trajectory 
in $(x,y)$ plane evolving up to time $t=300$
with initial energy as in left panels 
$E=1.5; 3; 10$ in top, 
center and bottom panels respectively.
In the top right panel an example of canonical 
variables (gray color) is shown with 
$\varphi=-0.75\pi$ (green/gray lines) and 
$\kappa=\sin{\pi/18}$ (violet/black lines), 
these variables are represented in the top left panel by 
the orange (gray) point.
Disk border is shown in right panels by red (gray) circle.
}
\end{figure}

It is convenient to describe the classical dynamics on
the Poincar\'e section using the canonical variables 
at the moment of the bounce with the disk.
We take the phase $\varphi$, given by the 
angle measured from $x$-axis, and
the conjugated dimensionless orbital momentum $\kappa=\sin{\theta}$, 
where $\theta$ is the angle of 
momentum $\vec{p}$ counted from the normal to the circle
(see Fig.~\ref{fig1}). Such a pair of conjugated variables 
represents a standard choice for the description of  
dynamics in billiards (see e.g. \cite{haake,stockmann}).
 
Figure \ref{fig1} shows that 
almost all phase space, accessible at a given energy, is chaotic 
(see e.g. \cite{chirikov,lichtenberg} on properties of dynamical chaos).
Only very tiny isands of regular 
motion are found at $E=3$ (practically not visible 
on the Poincar\'e section). At the energy 
$E=1.5$ the dynamics exists only on one side of the disk 
(variation of $\varphi$ is bounded, $-0.4 < \varphi/\pi \leq 1$)
due to symmetry breaking of the system.
At larger energies the trajectories make complete 
rotations around the disk. The amplitude of
oscillations grows with energy
approximately in the same way as in a usual 2D oscillator 
in absence of disk. However, the scattering on disk
makes the dynamics chaotic in a similar manner as for a standard
Sinai billiard \cite{sinai1970}. 

\begin{figure}[t]
\begin{center}
\includegraphics[width=0.48\textwidth]{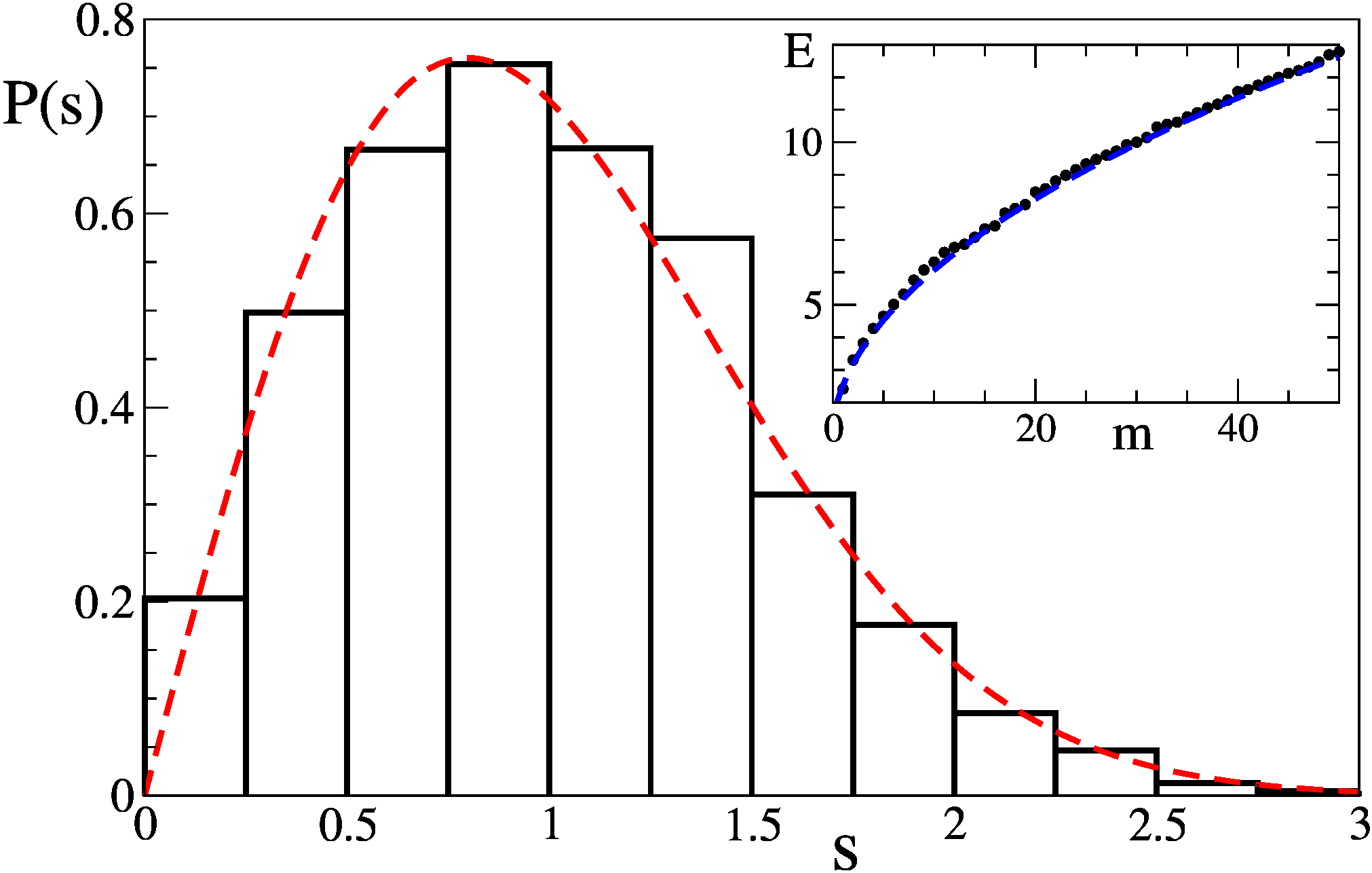}
\end{center}
\vglue -0.3cm
\caption{\label{fig2}
(Color online) Nearest-neighbor spacing distribution
$P(s)$ for the first 2500 unfolded eigenenergies of the 
Sinai oscillator (\ref{eq1}).
The red dashed curve represents the Wigner 
surmise $P(s)=(\pi s/2) \exp(-\pi s^2/4)$.
Insert panel shows energy
eigenvalues $E=E_m$ as a function of $m$ for the first 50 eigenvalues
$1\leq m \leq 50$.
Dashed blue curve represents the theoretical 
Weyl law $m(E)=E^2/(2\sqrt{2})-E/2$.
}
\end{figure}

The BEC evolution in the Sinai oscillator trap
is described by the GPE, which reads:
\begin{eqnarray}
\label{eq2}
 i\hbar{\partial\psi(\vec{r},t)\over\partial t} & = & -{\hbar^2\over 2m}
 \overrightarrow{\nabla}^2\psi(\vec{r},t) \\
\nonumber
 & + &
\left[ \frac{m}{2}(\omega_x^2 x^2+\omega_y^2 y^2) + V_d(x,y) \right] \psi(\vec{r},t) \\
\nonumber
& + &  \beta \vert\psi(\vec{r},t)\vert^2\psi(\vec{r},t) \; .
\end{eqnarray}
Here in  (\ref{eq2}), we use the same oscillator and disk parameters as in (\ref{eq1})
and take $\hbar=1$.  The wave function is normalized to unity
$W=\int |\psi(x,y)|^2 dx dy =1 $. 
Then the parameter $\beta$ describes the
nonlinear interactions of atoms in BEC.
All the results presented in the paper are expressed in these dimensionless units.
Thus the energy $E$ is expressed in units of $E_u=\hbar^2/(m {r_d}^2)=1$;
the distance is measured in units of $r_d=1$; time is measured in units
of $t_u=\hbar/E_u=1$; $\beta$ is measured in units of $\beta_u=1/{r_d}^2=1$.

Since the classical dynamics is chaotic and the measure of integrable
islands is very small it is natural to expect that at zero nonlinearity
$\beta=0$ the Sinai oscillator belongs to systems of quantum chaos 
\cite{bohigas,haake,stockmann}. Indeed, using the advanced methods
of quantum chaos on numerical computation of eigenenergies and eigenstates
in chaotic billiards (see e.g. \cite{vergini}), we find numerically
several thousands of eigenenergies $E_m$ and eigenstates $\phi_m$ (linear modes)
at $\beta=0$. 

The level spacing statistics $P(s)$ for the first $2500$ energy levels 
with the unfolding procedure (see e.g. \cite{haake}) is shown in 
Fig.~\ref{fig2}. The results are in good agreement with the Wigner
surmise confirming the validity of the Bohigas-Giannoni-Schmit conjecture
\cite{bohigas,haake}. 
The system energy $E_m$ grows with the level number $m$ 
in agreement with the Weyl law
as it is shown in the insert of Fig.~\ref{fig2}.

The linear eigenstates $\phi_m$ have a rather complex structure 
covering the accessible area in $(x,y)$ plane with 
chaotic fluctuations. The probability distributions
in the $(x,y)$ plane are shown for first $100$ eigenstates
in \cite{sinaiwebpage}; in addition, some eigenstates 
are also shown below.

The GPE (\ref{eq2}) can be also rewritten in the basis
of linear eigenstates $\phi_m$ using the completeness of this basis
and presenting the wave function by the expansion
$\psi(x,y,t)= \sum_m C_m(t) \phi_m(x,y)$,
where $C_m(t)$ are time dependent probability amplitudes in this basis.
Then in this basis the GPE reads:
\begin{equation}
i {{\partial C_m} \over {\partial {t}}}
=E_m C_m
+ \beta \sum_{{m_1}{m_2}{m_3}}
U_{m{m_1}{m_2}{m_3}}
C_{m_1}C^*_{m_2}C_{m_3} \; .
\label{eq3}
\end{equation}
Here the transitions
between eigenmodes appear only due to the nonlinear
 $\beta$-term and the transition matrix elements are
\begin{equation}
U_{m{m_1}{m_2}{m_3}} =
\int dx dy \, {\phi_m^*} \phi_{m_1} {\phi_{m_2}^*}  \phi_{m_3} \; .
\label{eq4}
\end{equation}
In our case, in absence of a magnetic field, the eigenstates
$\phi_m$ are real, but we keep the general expression
valid also for complex eigenstates.

A similar type of representation (\ref{eq4}) was used for the analysis of
effects of nonlinearity on the Anderson localization
in disordered lattices \cite{dls1993} known 
also as the DANSE model \cite{pikovsky,garcia}.
In this model it was found that a moderate nonlinearity 
leads to a destruction of the Anderson localization of 
linear eigenmodes and a subdiffusive spreading of 
wave packet over lattice sites with time.
Such a spreading has been studied by different groups
(see e.g. \cite{dls1993,pikovsky,garcia,fishman,flach}
and Refs. therein). Even if the representations for the GPE Sinai
oscillator (\ref{eq4}) and DANSE models are similar
there are significant differences:
(a) in DANSE the eigenenergies are bounded in a finite
energy band while here $E_m \propto \sqrt{m}$ are growing with $m$
(we note that for the Bunimovich billiard we have $E_m \propto m$
\cite{ermannepl}); (b) in DANSE the transitions $U_{m{m_1}{m_2}{m_3}}$
give coupling mainly between states inside 
the same localization length while here there are
transitions even between
very different $m$ values. At the same time 
we should say that the properties of matrix elements
 $U_{m{m_1}{m_2}{m_3}}$ are still waiting their detailed analysis
for the cases of Sinai oscillator and Bunimovich billiard.

We also note that the question of 
energy transfer to high energy modes
has certain links with the Kolmogorov turbulence
which is based on the concept of energy flow
from large to small scales via the inertial interval
(see \cite{zakharov,nazarenko} and Refs. therein).
The energy is injected at large scale
and absorbed on small scales and a presence of some small 
noise is assumed to induce thermalization.
In such an approach the ``quantum'' turbulence in 
GPE (or nonlinear Sch\"odinger equation (NLS))
has been studied in a rectangular 2D billiard 
\cite{nazarenko2014} and in 3D cube \cite{tsubota2015}. 
Due to the billiard shape chosen there, the ray dynamics
is integrable and it is not obvious if the dynamical thermalization
takes place in such a billiard in absence of noise.
Below we will see that 
at moderate nonlinearity and absence of noise
there is no dynamical
thermalization in  oscillator trap and
billiards of rectangular shape 
(see the later case in \cite{ermannepl}).
In fact, in purely dynamical systems (without external noise)
it is possible that 
the Kolmogorov flow to
high modes can be stopped by KAM integrability
and Anderson localization  \cite{dlskolm}.

Our results show that the quantum chaos for linear
eigenmodes facilitates onset of dynamic thermalization,
appearing in an isolated system without any noise,
when the strength of nonlinear term is above 
a certain dynamical thermalization border $\beta > \beta_c$.

\section{Thermodynamic formalism}

The dynamical thermalization in nonlinear chains
with disorder has been studied in \cite{mulansky,ermannnjp}
where it was shown that the quantum Gibbs distribution 
appears in an isolated system above a certain border 
of nonlinearity $\beta > \beta_c$. 
The dynamical thermalization for the GPE in a 
chaotic Bunimovich billiard has been established in \cite{ermannepl}.

Below for a reader convenience we present the thermalization formalism 
which directly follows from the standard 
statistical description of Bose gas \cite{landau}.
Indeed, we assume that the nonlinearity is moderate
and that the nonlinear term provides a small
energy shift which can be neglected. Then the energy levels are
those of the quantum Sinai oscillator with the usual 
quantum chaos properties and the energy levels $E_m$ at $\beta=0$.
As the energy and the norm of the system are conserved for
the quantum evolution of the system
 (Eq. \ref{eq2}),
the thermalization ansatz gives the 
steady-state probabilities $\rho_m$ on energy levels:
\begin{equation}
\label{eq5}
 \rho_m=1/[\exp[(E_m-E_g-\mu)/T]-1] \; ,
\end{equation}
where where $E_g = 1.685$ is the energy of the ground state,
$T$ is the temperature of the system,
$\mu(T)$ is the chemical potential dependent on temperature.
The parameters $T$ and $\mu$ are determined
by the norm conservation
$\sum_{m=1}^{\infty} \rho_m =1$ (we have only one particle in the system)
and the initial energy $\sum_m E_m \rho_m =E$.
The entropy $S$ of the system is determined by
the usual relation \cite{landau}: $S= - \sum_m \rho_m \ln \rho_m$.
The relation (\ref{eq5}), with normalization condition
and equation of energy,
determines the implicit dependencies on temperature
$E(T)$, $S(T)$, $\mu(T)$.

\begin{figure}[t]
\begin{center}
\includegraphics[width=0.48\textwidth]{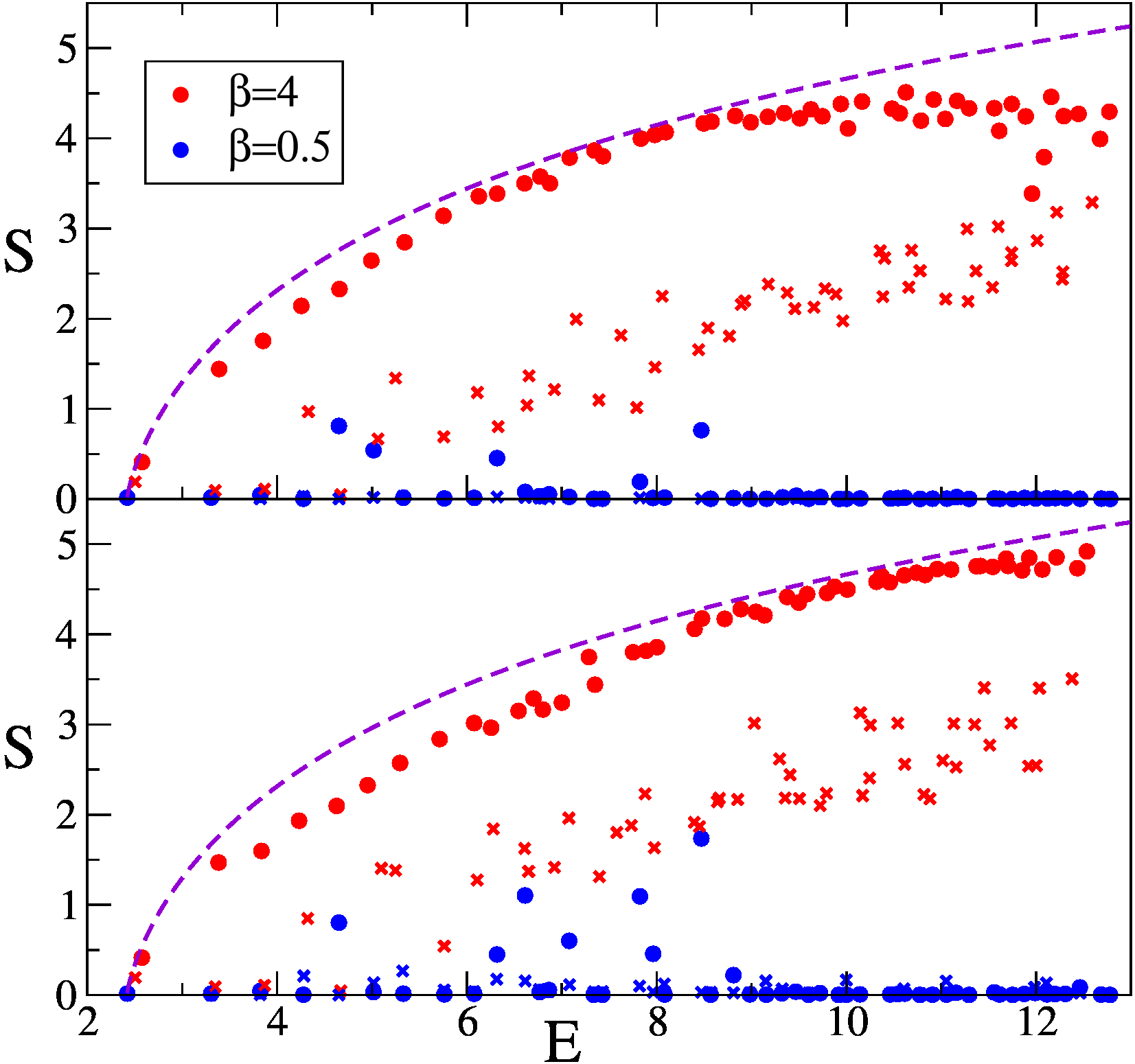}
\end{center}
\vglue -0.3cm
\caption{\label{fig3}
(Color online) Dependence of entropy $S$ on energy $E$, 
obtained from the GPE time evolution (\ref{eq2}) for initial states 
taken as first 50 eigenstates $\phi_m$ of the quantum Sinai oscillator. 
Blue (black) and
red (gray) symbols show the cases 
of nonlinearity $\beta=0.5$ and $\beta=4$ respectively, 
while circles and crosses represent the system with and 
without the elastic disk $r_d$ respectively.
The entropy $S$ is computed from $\rho_m=\langle |\langle m \vert\psi \rangle|^2 \rangle_t$ 
averaged over time intervals $t\in[500,1500]$ (top panel), and 
$t\in[1500,2500]$ (bottom panel). The dashed curve shows the theoretical
thermalization ansatz of Bose-Einstein distribution (\ref{eq5}).
}
\end{figure}

As it is pointed in \cite{mulansky,ermannnjp,ermannepl},
the advantage of  energy $E$  and entropy $S$ 
is that both are extensive variables,
thus they are self-averaging and due to that 
they have reduced fluctuations.
Due to this feature $S$ and $E$ are especially convenient for
verification of the thermalization  ansatz (\ref{eq5})
which gives the theoretical dependence $S(E)$ 
in the assumption that the dynamical thermalization
emerges in the GPE Sinai oscillator
due to dynamical chaos in absence of any external noise
or thermostat.

\section{Numerical results}

The numerical integration of GPE (\ref{eq2}) 
follows the approach used for the Bunimovich billiard in \cite{ermannepl}:
we introduce a space grid with size
$N_{s} =  n_x\times n_y=201 \times 141 = 28.341$ spacial points.
The time step is performed with the Trotter decomposition 
of linear and nonlinear terms with a time step
$\Delta t =0.01$. Thus, the nonlinear term gives the 
wave function transformation in coordinate space
${\bar \psi(x,y)} = \exp(-i\Delta t \beta |\psi(x,y)|^2)  \psi(x,y)$,
which is then transformed from coordinate space to the linear eigenbasis $\phi_m$
(we use $N_e=2000$ linear eigenstates).
The transformation from space grid to linear eigenfuntion index $m$ is done
via a precomputed transfer matrix $A_{jm}$ 
(here $1 \leq j \leq N_s$ is an index of space grid). After that the linear
propagation step is performed with expansion coefficients 
in the eigenstate basis
$C(t+\Delta t) =\exp(-i \Delta t E_m) C_m(t)$. 
Then  the back transformation from linear basis $\phi_m$ to 
coordinate basis finally gives $\psi(t+\Delta t)$.
In addition to the space representation $\psi(x,y,t)$ we also compute the
wave function in the momentum representation
using the standard relation
$\phi(\vec{p},t)=
\int \psi(\vec{r},t) \exp(-i \vec{r} \vec{p}/ \hbar) dr^2 / {2\pi\hbar}$
with a discrete Fourier transform.
The time evolution computed in this way
gives an approximate energy $E$ and norm $W$ conservation.
Typically we have at time $t=2000$ the variation of 
these integrals being $\delta W /W =0.001 (0.004)$
and $\delta E/E =0.002 (0.004)$ for initial state at $m=10 (40)$
respectively. We return to a more detailed discussion
of the accuracy of computations in Section V.

\begin{figure}[t]
\begin{center}
\includegraphics[width=0.48\textwidth]{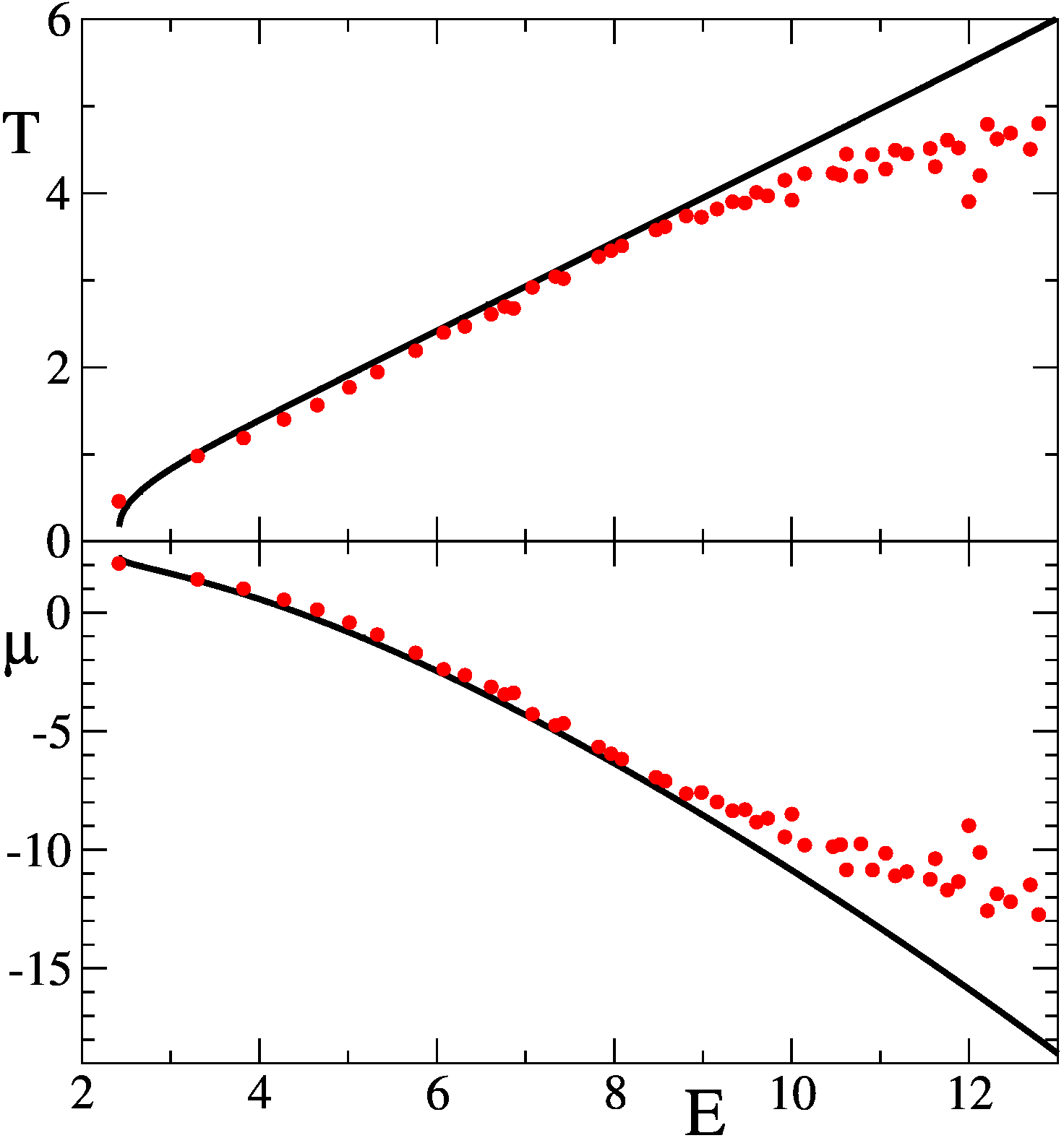}
\end{center}
\vglue -0.3cm
\caption{\label{fig4}
(Color online) Dependence of temperature $T$ and 
chemical potential $\mu$ on 
energy shown in top and bottom panels respectively.
Black curves represent the theoretical ansatz given 
by the Bose-Einstein distribution (\ref{eq5}), while
red circles represent numerical data 
 $T=(T_1(E)+T_2(S))/2$ and $\mu=(\mu_1(E)+\mu_2(S))/2$ 
($T_{1,2}$ and $\mu_{1,2}$ values are
computed from $E$ and $S$ respectively)
for initial states given by first 50 linear eigenstates 
and probabilities
$\rho_m=\langle |\langle m \vert\psi \rangle|^2 \rangle_t$ 
averaged over interval $t\in[500,1500]$. Here $\beta=4$.
}
\end{figure}

During the time evolution we determine  the probabilities
of wave function in the linear eigenmodes 
$\rho_m=\langle |\langle m \vert\psi \rangle|^2 \rangle_t = 
\langle  |C_m|^2| \rangle_t$ 
averaged over a certain time interval. 
Usually we choose this time interval as  approximately last half (or similar to that) 
of the whole evolution range to obtain approximate steady-state values of $\rho_m$.
From these averaged values we determine the entropy
$S= - \sum_m \rho_m \ln \rho_m$. Thus starting from 
different initial states, chosen as linear eigenstates $\psi_m$,
we obtain numerically the dependence $S(E)$ which is compared with the
prediction of the Bose-Einstein thermalization ansatz (\ref{eq5}).

\begin{figure}[h]
  \includegraphics[width=0.47\textwidth]{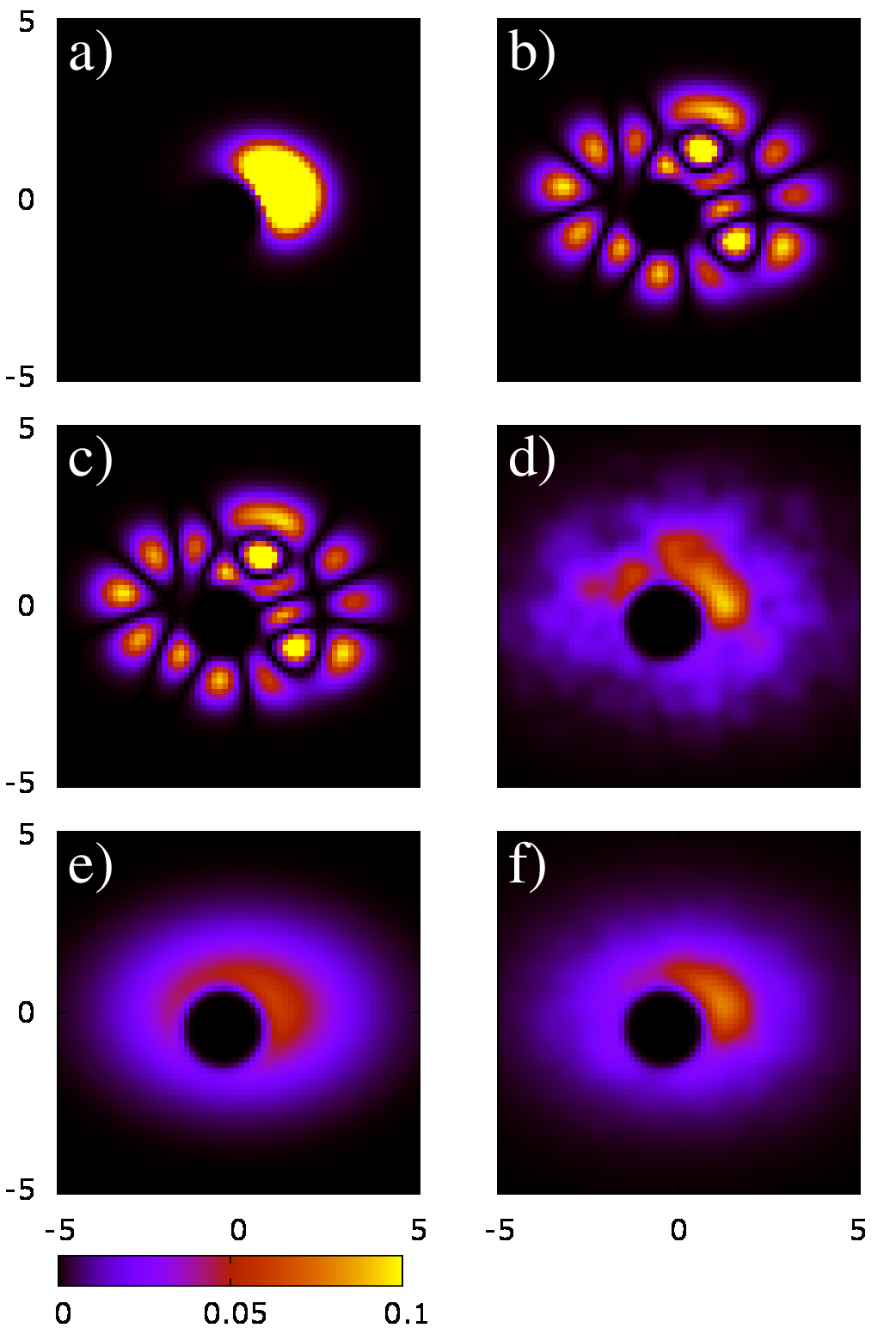}
 \caption{(Color online) Spacial probability distributions
$|\psi(x,y)|^2$ for the GPE Sinai oscillator.
Panels a) and b) show the linear eigenstates $m=1$ (ground state) 
and $m=24$  with eigenergies $E_1=2.417$ and $E_{24}=9.16$
respectively. Panels  c) and e) have 
the initial state $m=24$ of panel b) and  
show the average distributions at long times
with averaging over large interval 
$t\in[1500,2500]$ for $\beta=0.5$ and $\beta=4$ respectively. 
Panel d) shows  
the average distribution for short time interval (snapshot) 
$t\in[2000,2005]$ for $\beta=4$.
Panel f) shows the thermal
Bose-Einstein distribution (\ref{eq5}) for energy $E_{24}=9.16$
(to be compared with panel e)).
Probability is shown by color bar changing from zero (black) to
maximum (yellow/gray). Numbers in horizontal and vertical 
axes show the scales in $x$ and $y$ respectively.  
}
 \label{fig5}
\end{figure}

The comparison of numerical data with the theoretical curve obtained from
(\ref{eq5}) is shown in Fig.~\ref{fig3}. It is clear that 
for small $\beta =0.5$ the nonlinear term leads to excitation of certain
eignemodes of the Sinai oscillator but the numerical data
for $S(E)$ are pretty far from the theoretical dashed curve 
given by the Bose-Einstein distribution (\ref{eq5}).
For the 2D oscillator without disk the excitation
is significantly weaker than for the Sinai oscillator
with  $S(E)$ values being very far form the theory
both for $\beta=0.5, 4$. In contrast to that, for the 
GPE Sinai oscillator at $\beta=4$ our numerical data for $S(E)$
are close to the theoretical thermalization ansatz.
For the time interval $t\in[500,1500]$ (Fig.~\ref{fig3} top panel)
the obtained  $S$  values
for $35 < m \leq 50$ ($10.5 <E <13$ are somewhat below the 
theoretical curve. We attribute this to the fact
that for large $m$ the effective amplitude 
of nonlinear term in (\ref{eq2}) 
is reduced $|\psi|^2 \sim 1/x^2 \sim 1/E \sim 1/\sqrt{m}$
and hence, it takes a longer time for dynamical chaos to
establish dynamical thermalization.
Actually, the nonlinear energy shift as
$\delta E_\beta \sim \beta  |\psi|^2 \sim \beta/\sqrt{m}$
and therefore, the thermalization time $t_T$ should be
at least proportional to $t_T \sim 1/{\delta  E_\beta} \sim \sqrt{m}/\beta$.
Indeed, at large times 
$t\in[1500,2500]$ (Fig.3 bottom panel)
we find the values of $S$
being significantly more close to the thermalization ansatz
for  $35 < m \leq 50$ ($10.5 <E <13$).
Thus the results of Fig.~\ref{fig3} show the onset
of dynamical thermalization for moderate 
values of $\beta > \beta_c \sim 1$.

\begin{figure}[h]
  \includegraphics[width=0.47\textwidth]{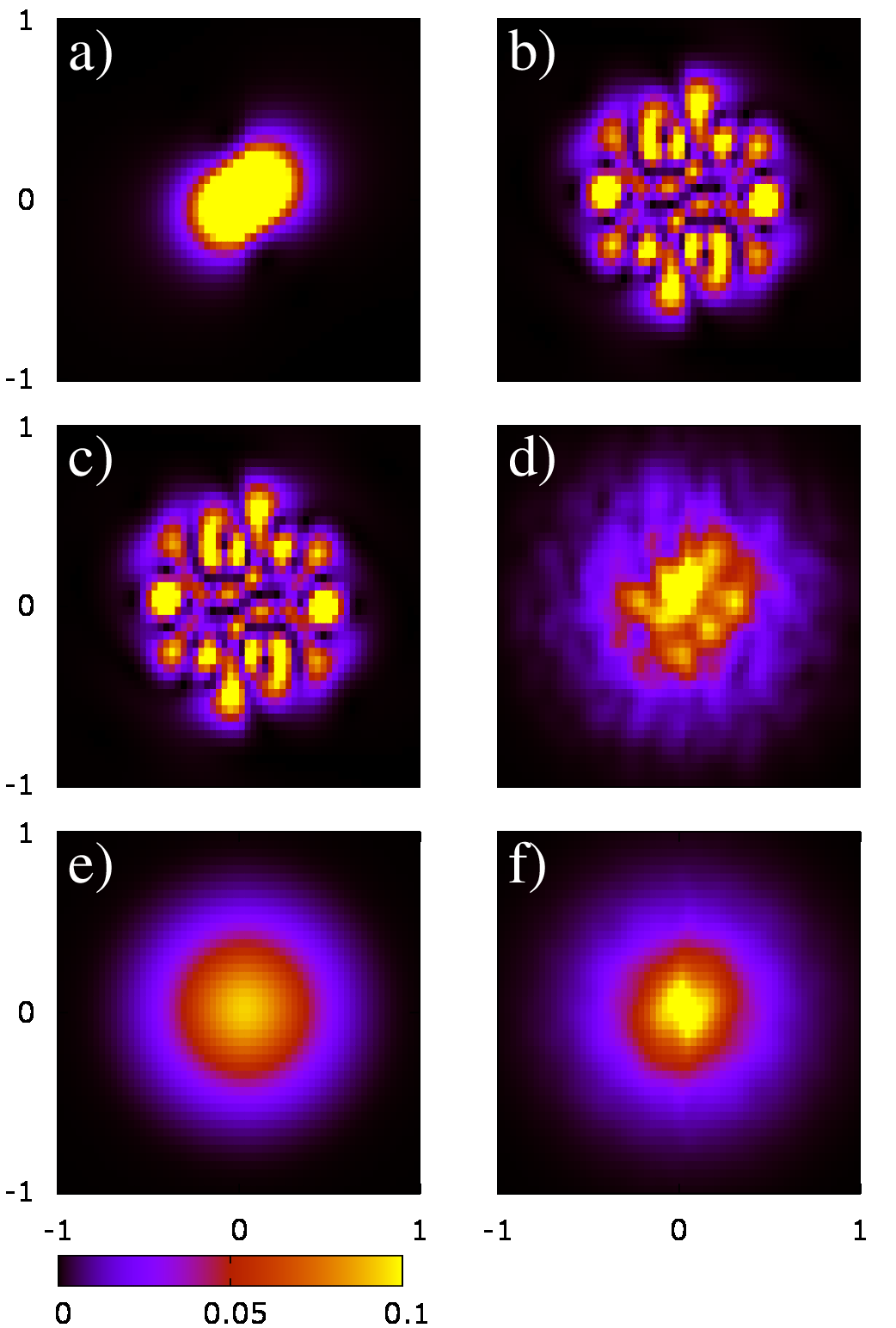}
 \caption{(Color online) Momentum probability distributions
$|\psi(p_x,p_y)|^2$ shown for the same cases as in 
6 panels of Fig.\ref{fig5}. 
Probability is shown by color bar changing from zero (black) to
maximum (yellow/gray). Numbers in horizontal and vertical 
axes show the scales in $x$ and $y$ respectively.
}
 \label{fig6}
\end{figure}

As for the case of Bunimovich stadium \cite{ermannepl},
we expect that the thermalizaiton border $\beta_c \sim 1 $
(definitely $0.5 < \beta_c <4$)
is independent of $m$. Indeed, the nonlinear energy shift
$\delta E_\beta \sim \beta/\sqrt{m}$ and the level spacing
$\Delta E \sim 1/\sqrt{m}$
scale with $m$  in a similar way
(see Fig.~\ref{fig2} for the dependence $E_m$) so that we can expect  
chaos and thermalization to be set in at $\delta E_\beta > \Delta E$,
thus leading to $\beta_c \sim 1$.
Indeed, similar estimates have been confirmed in
systems of coupled nonlinear  oscillators
\cite{dls1993,ermannnjp,chirikovdls}).

%
%
%
%
%
%

Another confirmation of the onset of dynamical thermalization is shown 
in Fig.~\ref{fig4}. According to (\ref{eq5})
the temperature can be determined from an initial
energy $E$, giving $T_1(E)$, or from an average
value of $S$, giving $T_2(S)$. In a similar way we can determine
$\mu_1(E)$ and $\mu_2(S)$. The dependence of average numerical values $T=(T_1+T_2)/2$,
$\mu=(\mu_1+\mu_2)/2$ on energy $E$ are shown in Fig.~\ref{fig4}
being in a good agreement with the thermalization ansatz (\ref{eq5}).
The observed deviations for $35 < m \leq 50$ 
are related with the lack of sufficiently large time evolution in the simulations . 

The transition from nontermalized (quasi-integrable) regime to
dynamical thermalization is also visible from the spacial probability
distributions shown in Fig.~\ref{fig5}.
We start from a typical initial state $m=24$ shown in panel (b).
For $\beta =0.5 < \beta_c$ a snapshot distribution at $t=2000$ (panel c))
remains very similar to the initial state showing the absence of thermalization
and dominance of the initial mode. In contrast to that
for $\beta =4 > \beta_c$ a snapshot at $t=2000$ (panel d)) shows 
that the distribution have a dominant component at the ground state mode
shown in panel a). The  distribution averaged over a large time interval,
assumed to be close to a steady-state, is shown in panel f).
It is indeed very similar to the theoretical steady state
probability distribution
$|\psi_{st}(x,y)|^2 = \sum_m \rho_m |\phi_m(x,y)|^2$
where $\rho_m$ are given by the Bose-Einstein distribution (\ref{eq5})
(see Fig.~\ref{fig5} panel e)).

The  probability distributions in the momentum space
$(p_x,p_y)$, corresponding to cases of Fig.~\ref{fig5}, 
are shown in Fig.~\ref{fig6}. These data also show a clear absence of thermalization 
for $\beta=0.5$ (panels b), c)) and  close similarity between 
the theoretical distribution (panel e)) and average distribution (panel f)).

\begin{figure}[h]
  \includegraphics[width=0.47\textwidth]{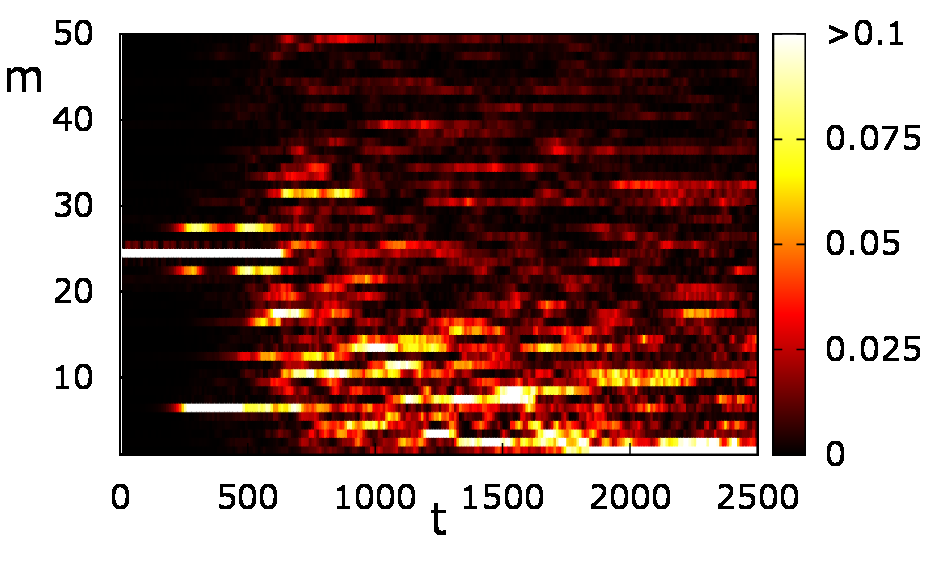}
 \caption{(Color online) Time evolution of probabilities $\rho_m(t)$
in the basis of linear eigenmodes for the initial state $m=24$ at $\beta=4$.
The probabilities $\rho_m(t)$ are averaged over time $\delta t=10$
to reduce fluctuations. Color bar shows probabilities from zero (black) to maximum
(white).
}
 \label{fig7}
\end{figure}

Thus the results of this Section provide a good confirmation
of onset of dynamical thermalization at
moderate nonlinearity $\beta \approx 4 > \beta_c \sim 1$.
However, it is also important to analyze the larger scale evolution on times 
being larger than those considered here with $t \leq 2500$.
This consideration is presented in next Section.

\section{Large time scales and numerical methods}

The question about long time scale evolution of Eq.~\ref{eq2} 
requires further extensive studies with improved
accuracy on numerical simulations.
Indeed, at times $t > 2500$ we see a tendency of 
probability accumulation at the ground state $\phi_1$
of linear system. First signs of this trend
are seen in Fig.~\ref{fig7}  showing evolution $\rho_m(t)$
with appearance of large values of $\rho_1$ at $t \approx 2300$. 

\begin{figure}[h]
  \includegraphics[width=0.47\textwidth]{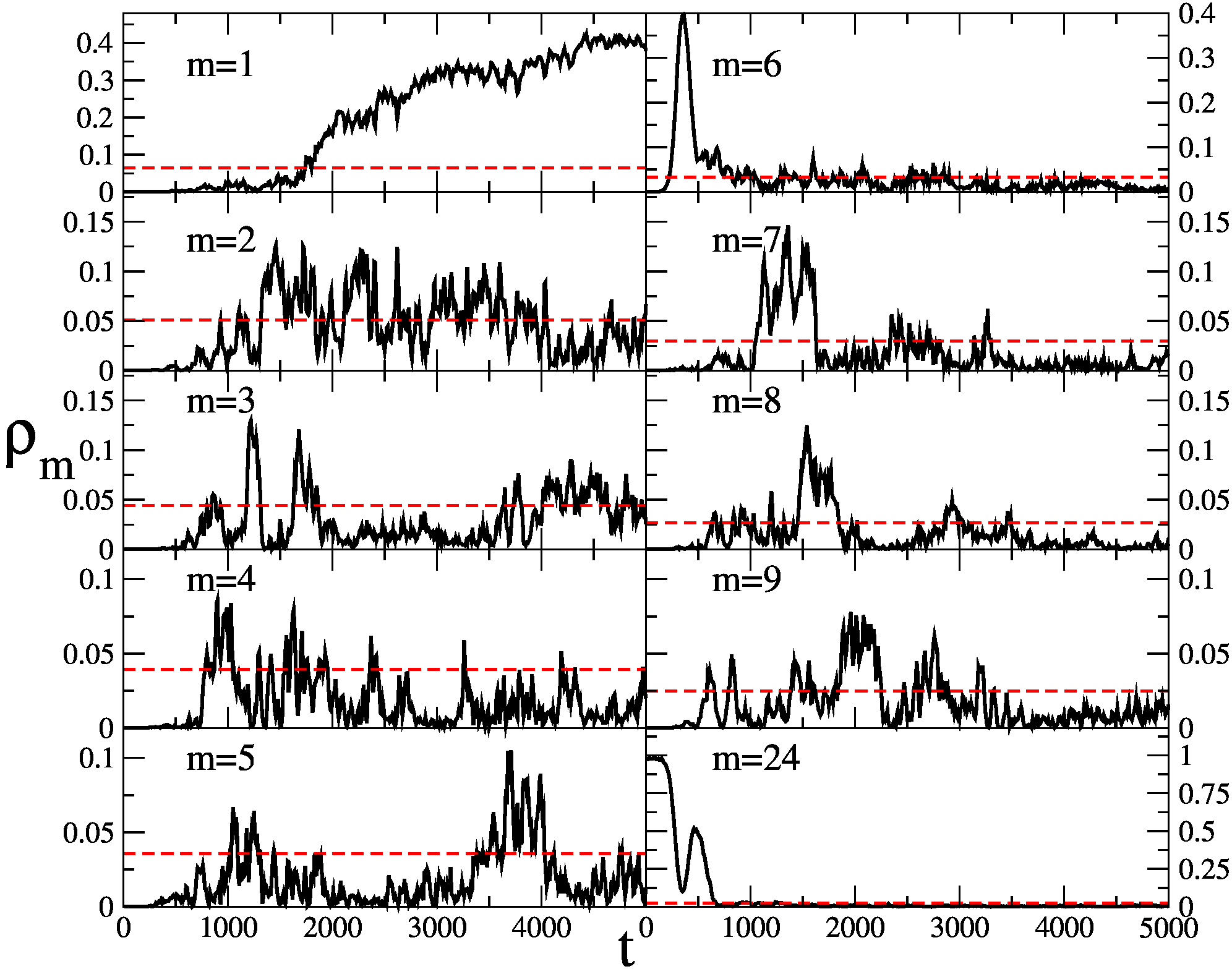}
 \caption{(Color online) Time evolution of probabilities $\rho_m(t)$
in the basis f of linear eigenmodes for the initial state $m=24$ at $\beta=4$.
The probabilities $\rho_m(t)$ are averaged over time $\delta t=10$
to reduce fluctuations. Ten panels show $\rho_m(t)$ 
for $m=1,2,3,4,5,6,7,8,9$ and $m=24$; red dashed lines show
the theoretical values of $\rho_m$ from the thermalization ansatz 
of Bose-Einstein distribution (\ref{eq5}). 
}
 \label{fig8}
\end{figure}

\begin{figure}[h]
  \includegraphics[width=0.47\textwidth]{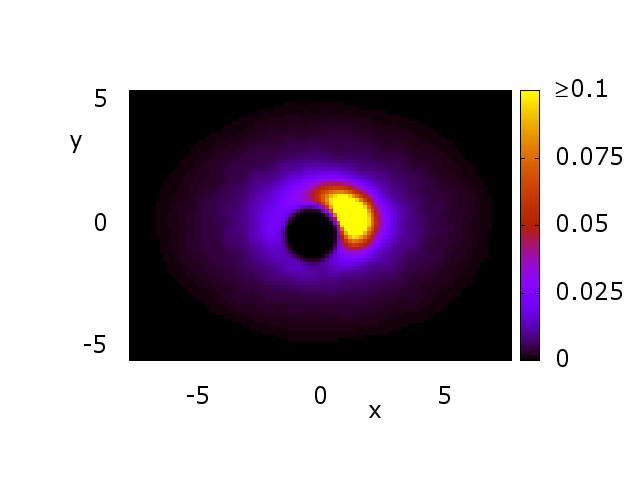}
 \caption{(Color online) Long time average 
probability distribution $<|\psi(x,y)|^2>_t$ shown in coordinate space.
The  evolution starts from the initial 
linear eigenstate $m=24$ (see Fig.~\ref{fig5}), 
the average is done over the time interval $t\in[4000,5000]$;
here $\beta=4$ (to compare with panels of Fig.~\ref{fig5}
with same colors). Numbers in horizontal and vertical 
axes show the scales in $x$ and $y$ respectively.
 }
 \label{fig9}
\end{figure}

A more detailed view of dependence of $\rho_m(t)$ on time $t$ for several
selected $m$ values is presented in Fig.~\ref{fig8} up to $t=5000$.
For $m>1$ there are large fluctuations in time
and it is clear that averaging on large time scales is required to obtain
statistically stable values of $\rho_m$. For the extensive variables
like energy $E$ and entropy $S$ these fluctuations are reduced and 
that is the reason due to which the data for the curve $S(E)$ are less
fluctuating. However, for $m=1$ in Fig.~\ref{fig8} there is a steady growth
with a the apparent saturation at $t \approx 4500$ at the value $\rho_1 \approx 0.4$ which is
by a factor $10$ larger than the theoretical value shown by the dashed line.
The spacial distribution of probability at such large times also 
demonstrates a strong accumulation of probability 
at the ground state as it is shown in Fig.~\ref{fig9}.
A video of evolution on large times is available at \cite{sinaiwebpage}.
A similar probability accumulation at the ground state
is seen for other initial states with $\psi(t=0)=\phi_{40}; \phi_{60}$.
We note that we had no such accumulation of probability
in the ground state in 
the studies of  GPE Bunimovich billiard \cite{ermannepl}.

\begin{figure}[h]
  \includegraphics[width=0.47\textwidth]{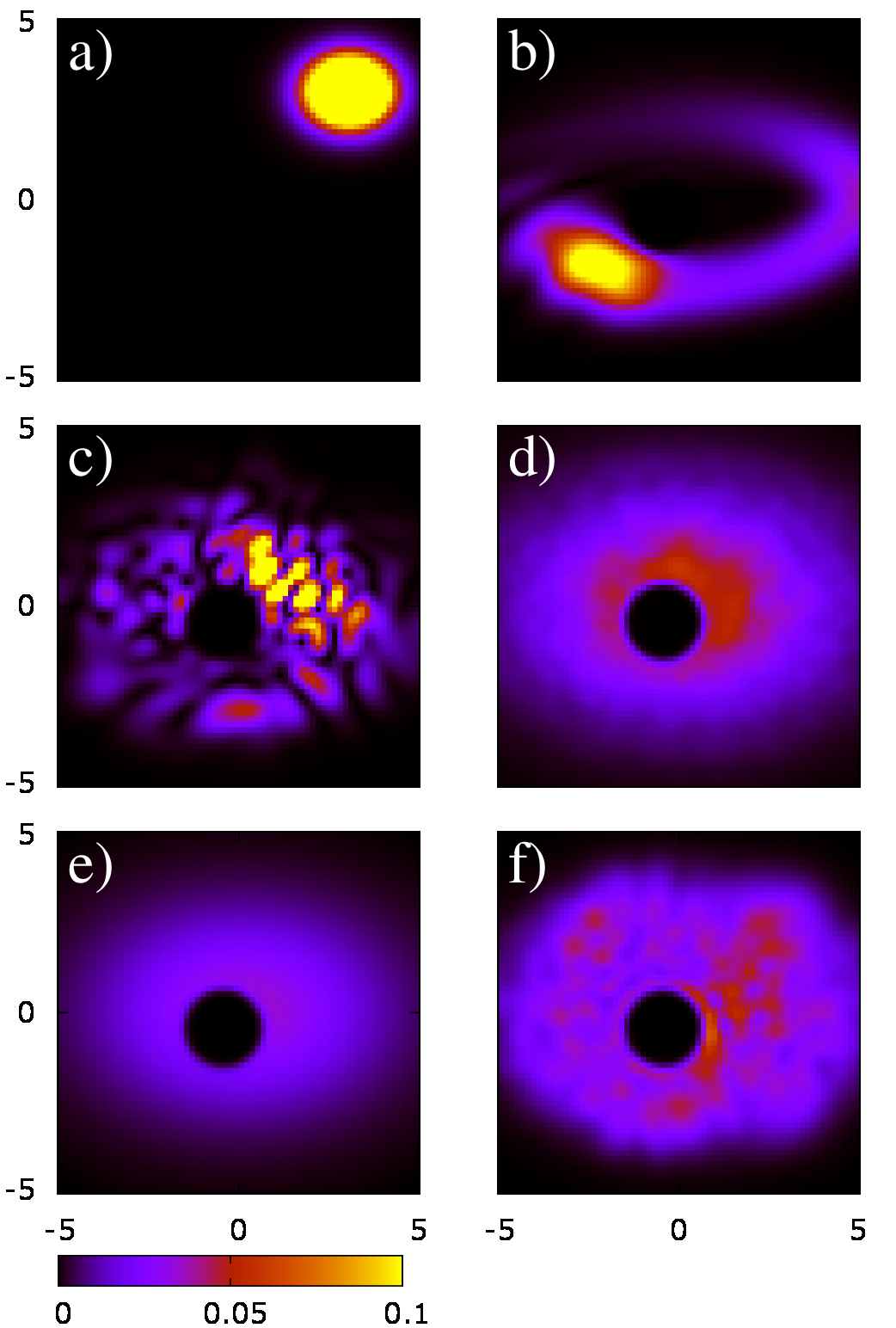}
 \caption{(Color online) 
Space probability distributions
$|\psi(x,y)|^2$ for the case of an initial state given by a 
coherent wave packet centered at $(x_0,y_0)=(3,3)$ 
(and zero probabilities on the disk).
Panel a) shows the initial state distribution with $E=14.707$;
panels b) and c) show distributions with $\beta=4$ 
at $t=2.9$ and $t=8$ respectively.
Long time averages over the interval $t\in[4000,5000]$ 
are shown in panel d) and f)
for $\beta=4$ and $\beta=0$ (linear evolution)  respectively.
Panel e) shows the theoretical space probability from the Bose-Einstein 
distribution (\ref{eq5}) for energy $E=14.707$.
Numbers in horizontal and vertical 
axes show the scales in $x$ and $y$ respectively.
}
 \label{fig10}
\end{figure}

We also considered the time evolution for a different 
initial state taken as initial coherent wave packet
centered at certain position $(x_0,y_0)$
(of course the packet is only approximately Gaussian
since the probability is zero at the disk border).
The initial distribution and snapshots at a few moments of time
are shown in Fig.~\ref{fig10} (panels a), b), c)).
The time evolution video is available at \cite{sinaiwebpage}
for $\beta=0$ and $\beta=4$ on short and long time scales.
The probability averaged over time interval   $t\in[4000,5000]$
is shown in panels  f) and d)  $\beta=0$ and $\beta=4$ respectively.
The case with interactions shows a tendency of
accumulation of probability at low $m$ modes 
in a qualitative agreement with the 
thermal ansatz distribution shown in panel e).
However, it is visible that the GPE case has a 
larger probability on low energy modes $m$.
In contrast, the case with $\beta=0$ (panel f))
has large probability in initial high modes $m$.

\begin{figure}[h]
  \includegraphics[width=0.47\textwidth]{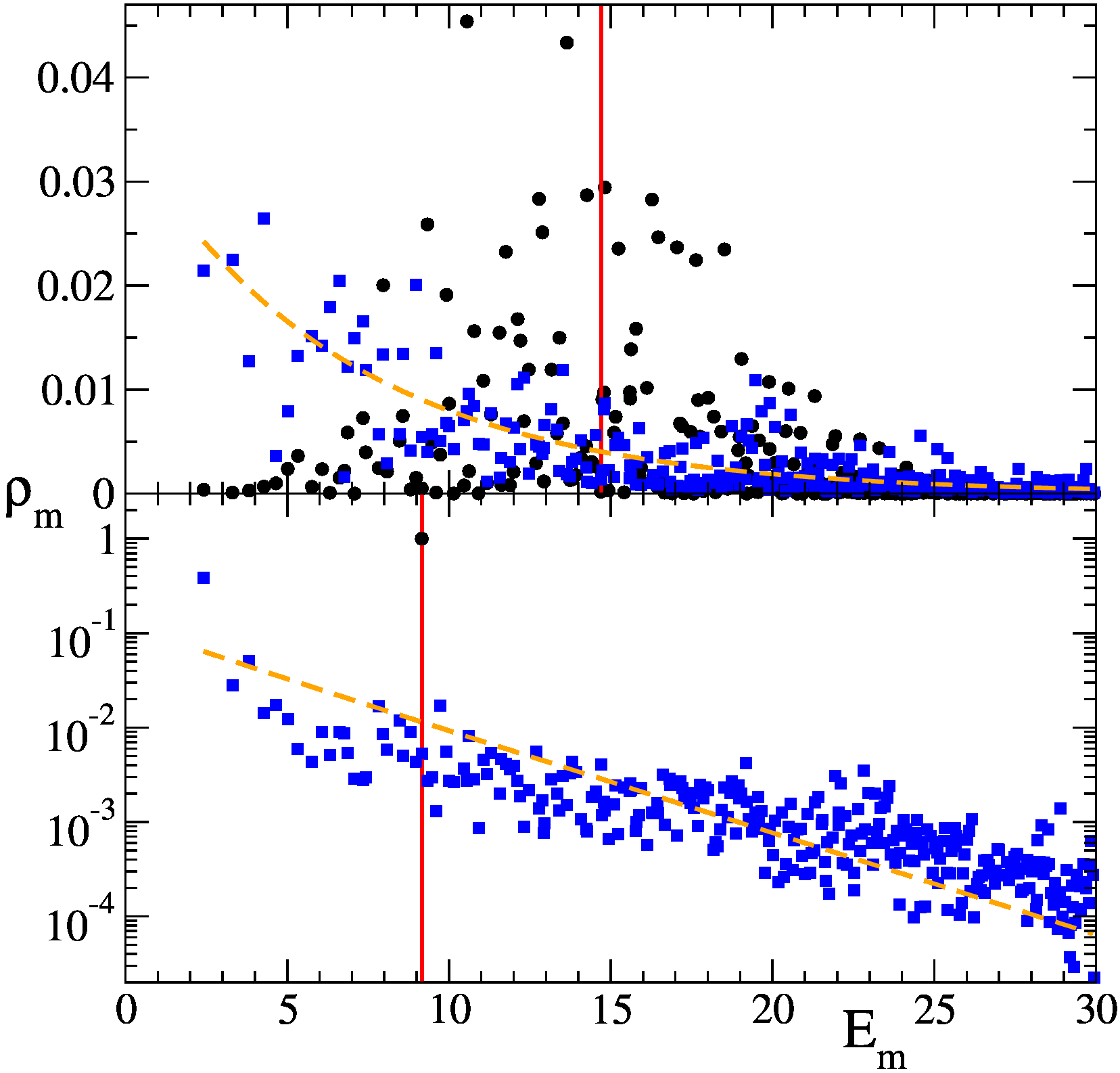}
 \caption{(Color online) Probability $\rho_m$ vs. energy of linear eigenstate $E_m$.
Top panel shows the case of initial state given 
by a coherent state localized at $(x_0,y_0)=(3,3)$ 
from Fig.\ref{fig10}; bottom panel shows the case of 
initial linear eigenstate $m=24$ (semi-logarithmic scale).
Black circles and blue squares show $\rho_m$ for initial state 
and for long time average with $\beta=4$ and $t\in[4000,5000]$ respectively. 
Solid red line shows the total energy, while dashed orange
line represents the theoretical distribution $\rho_m$ (\ref{eq5})
for corresponding energy $E=14.71$ and $E=9.16$ 
(top and bottom panels respectively).
 }
 \label{fig11}
\end{figure}

The average probability distributions over linear modes
are shown at large times in Fig.~\ref{fig11}.
For the initial state in a form of coherent packet 
there is a clear displacement of highest probabilities from initial
energies $E_m \approx 14.7$ to modes with $m=1,2,4$. In this case there is
no strong accumulation of probability at the ground state.
In fact, the average value of $\rho_1 \approx 0.22$ is comparable with the
thermalization ansatz value. 
Also for this state with energy $E \approx 14.7$
we find numerically the average entropy value
$S=4.96$ being close to the theoretical value of (\ref{eq5})
with $S=5.47$.
However, the fluctuations
of average $\rho_m$ probabilities are too strong and the comparison with 
the thermalization ansatz curve is only qualitative.
For the initial state with $\phi_{24}$ the probability $\rho_1$
is significantly larger then the theoretical value
(by a factor $10$),  then the decay of $\rho_m$ with $E_m$
approximately follows the thermalization ansatz
but also the fluctuations are large.
We think that 
the fluctuations are larger for the case of coherent state
due to a larger number of initially exited linear eigenmodes
compared to the case with $\psi(t=0)=\phi_{24}$.
Also the initial energy of the coherent state is $E=14.71$ being larger than
energy of $m=24$ with $E=9.16$ and hence longer times are required for complete
thermalization of the coherent state.

\begin{figure}[h]
  \includegraphics[width=0.47\textwidth]{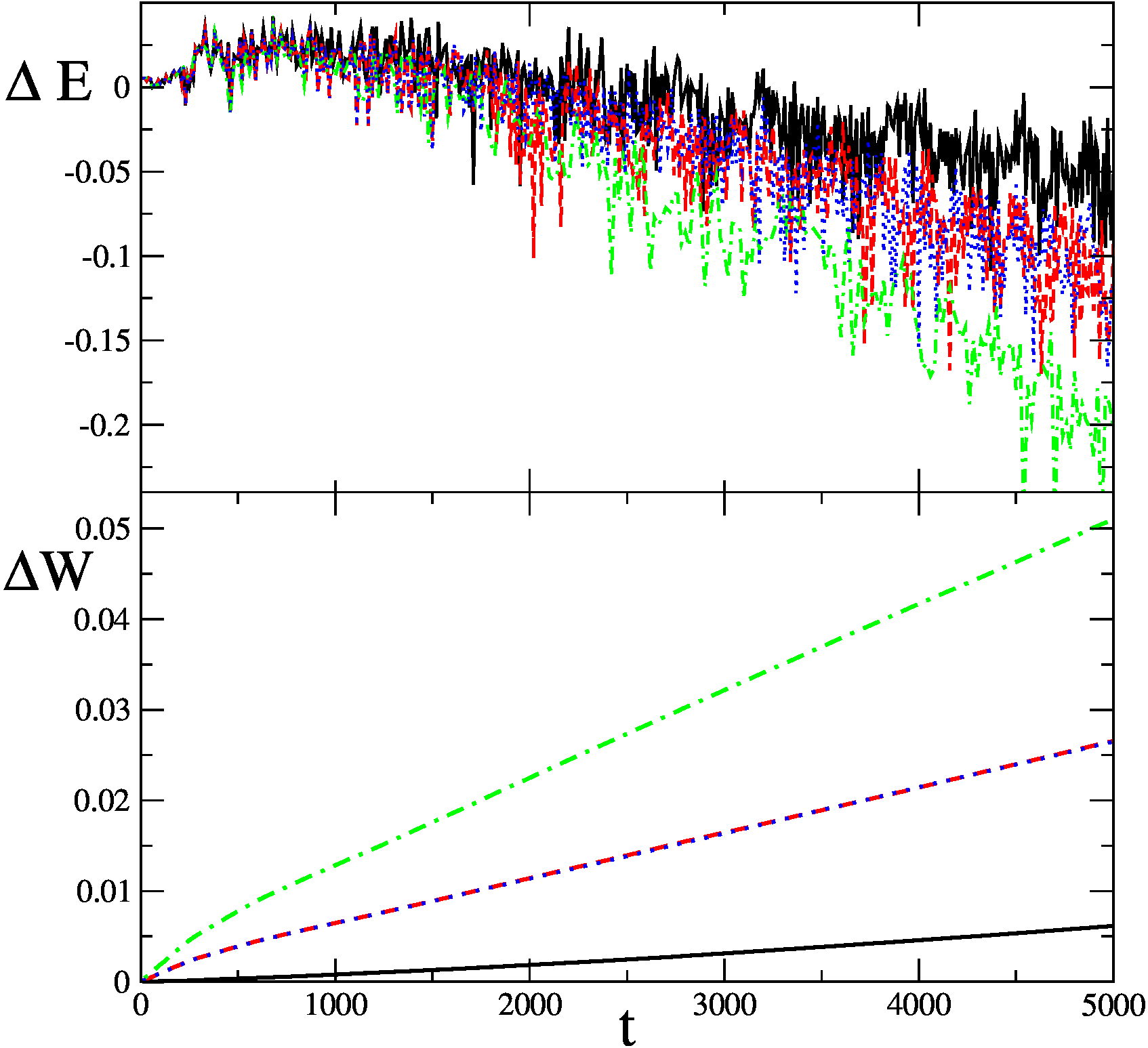}
 \caption{(Color online) Time evolution of  energy and norm conservation for 
different parameters of numerical simulations with
the initial linear eigenstate $m=24$ and $\beta=4$.
Top panel and bottom panel show the evolution of 
$\Delta E=E-E_{24}$ and $\Delta W=1-\sum_m\rho_m$.
The simulation parameters are: 
$N_e=2000, \Delta t=0.01$ and 
number of spacial lattice of $N_s=201\times 141$ points in the rectangle given 
by $(x_{\text{min}}, y_{\text{min}})=(-\sqrt{234} 
\approx -15.3,-\sqrt{117} \approx -10.82)$ and  
$(x_{\text{max}}, y_{\text{max}})=(\sqrt{234} 
\approx 15.3,\sqrt{117} \approx 10.82)$ 
(black solid curves);
$N_e=1000, \Delta t=0.01$ with  $N_s=201\times 141$ (red dashed curves);
$N_e=1000, \Delta t=0.005$ with $N_s=201\times 141$ (green dot-dashed curves);
$N_e=1000, \Delta t=0.01$ with  $N_s=361\times 255$ lattice points 
in the same region (blue dotted curves); 
red and blue curves practically coincide in the bottom panel.
  }
 \label{fig12}
\end{figure}

\begin{figure}[h]
  \includegraphics[width=0.47\textwidth]{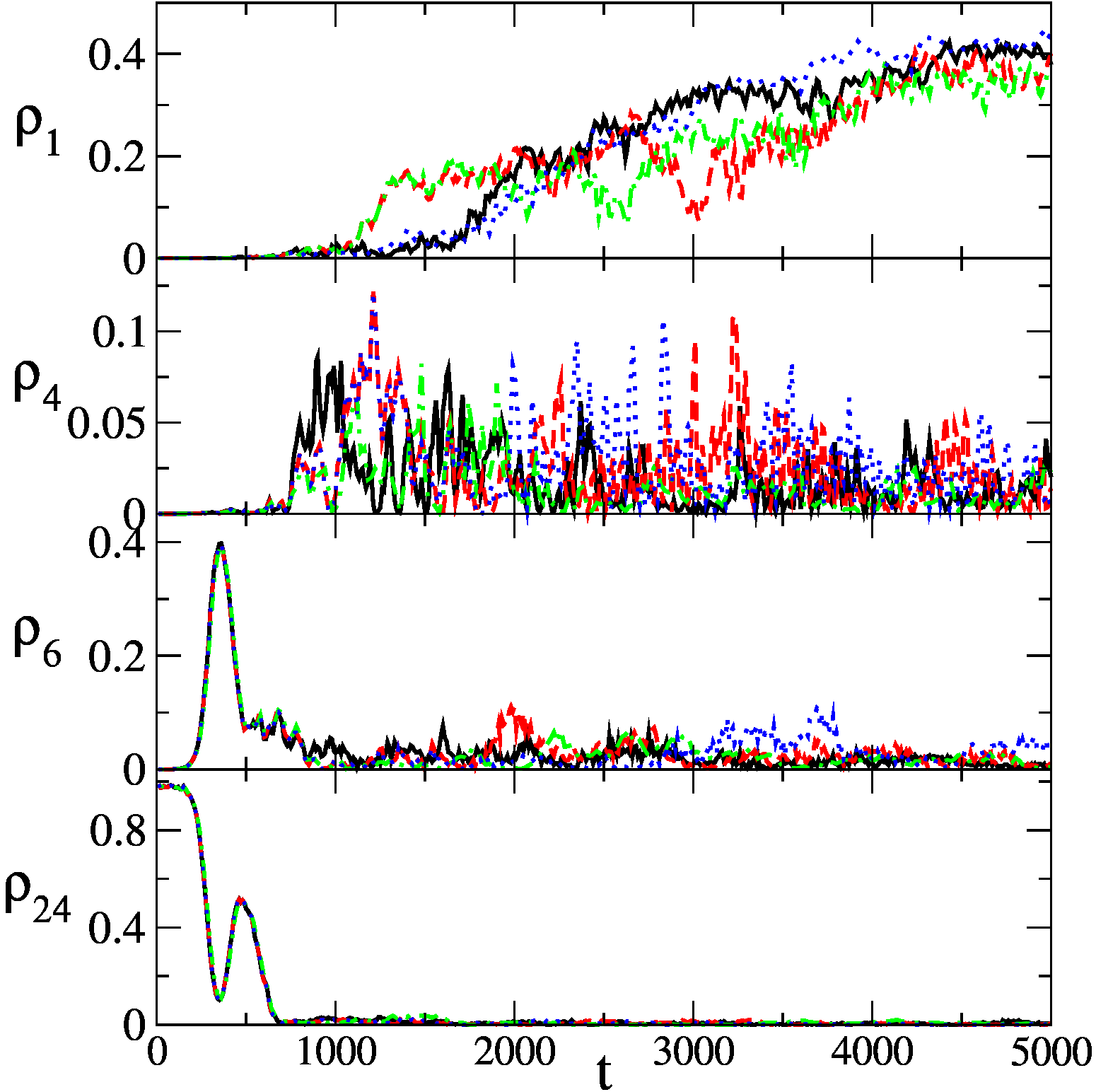}
   \caption{(Color online) Time evolution of probabilities $\rho_m(t)$
for parameters of Fig.~\ref{fig12} shown for $m=1, 4, 6, 24$
with the same colors as in Fig.~\ref{fig12}.
 }
 \label{fig13}
\end{figure}

To check the validity of  numerical integration for large time scales
we performed a number of checks shown in Fig.~\ref{fig12} and Fig.~\ref{fig13}. 
With this aim we varied the integration time step $\Delta t$,
the number of lattice points $N_s$ in the coordinate space 
$(-x_{min},x_{max}; -y_{min}, y_{max})$ and the spacial range
of the lattice determined by these min/max values of $x, y$.
The integration scheme gives a slow decrease of norm $W$ and energy $E$
with time indicating that there are some effective dissipation induced by
numerical integration.
These checks show that the most sensitive parameter is 
the number of eigenstates $N_e$ used in the transformation matrix
$A_{j,m}$ from coordinate space to linear eigenmodes.
Up to time $t \approx 800$ all numerical curves in Fig.~\ref{fig13}
give the same results showing the validity of numerical 
integration. However, for $N_e=1000$ the non-conservation of $W$ and $E$
becomes significant at large times giving significantly different
values of $\rho_1, \rho_4$ at $t \approx 1000 - 1500$ for $N_e=1000$ and 
$N_e=2000$. Thus we use the numerical parameters with $N_e=2000$
for all results presented in above Figs.~\ref{fig1}-~\ref{fig11}
since this choice provides the best conservation of $W$ and $E$.

The above verification show that our integration scheme
introduces a hidden intrinsic effective dissipation and due to this
reason we consider that this can be the reason of
probability accumulation at the ground state (or a few states 
with low $m$ values) at large times. Due to this reasons we
consider that our numerical results are reliable only up to
finite times $t \leq t_{num} \approx 2500$. 
We believe that better symplectic integration schemes should
be developed to extend numerical studies 
for larger time scales.

\begin{figure}[h]
  \includegraphics[width=0.47\textwidth]{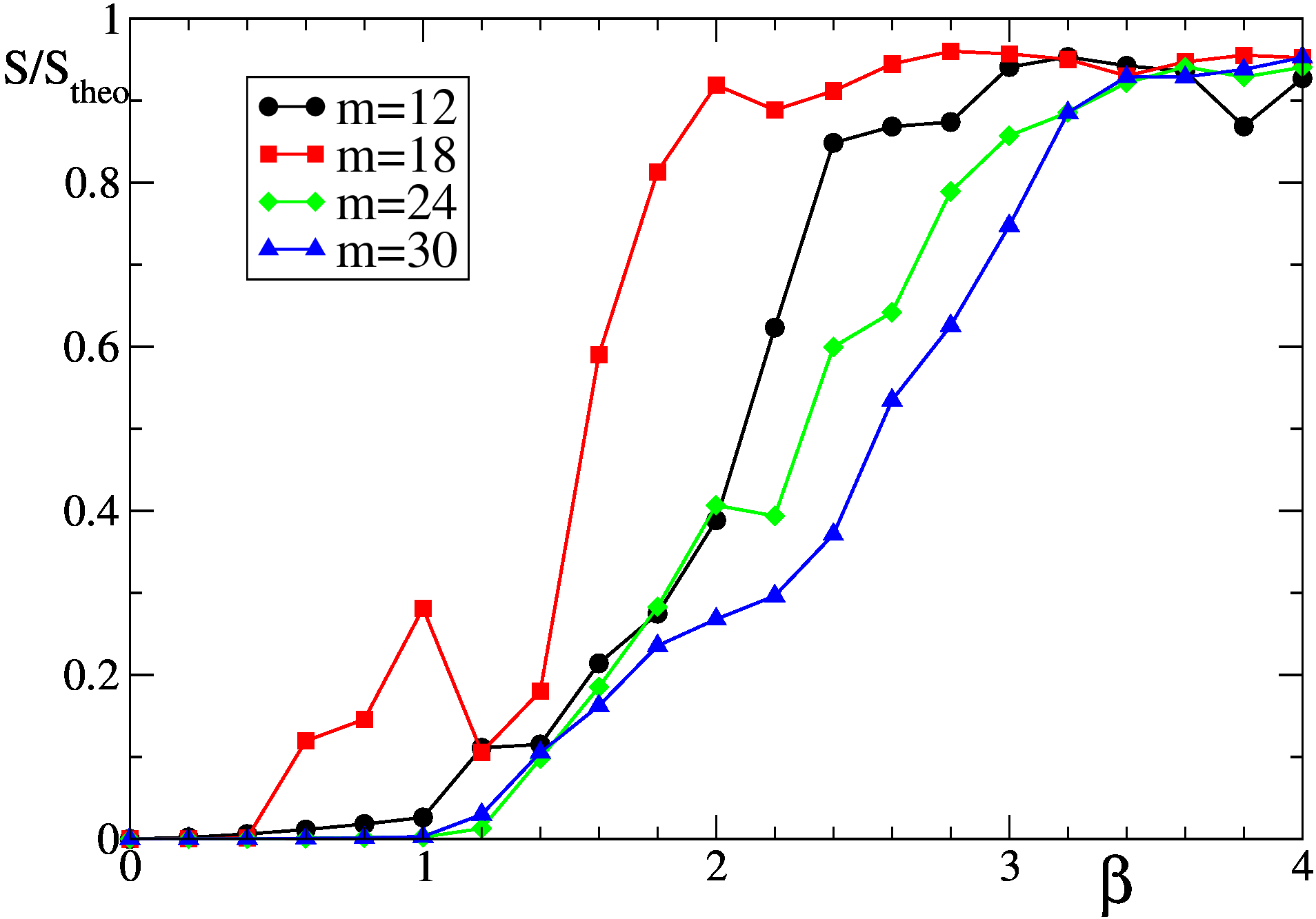}
   \caption{(Color online) Entropy $S/S_{theo}$ as a function of nonlinear
parameter $\beta$. The entropy $S$ is computed numerically
in linear basis with probability average $\rho_m$ over
the time interval $t\in[1000-2000]$.
Initial states are linear eigenstates with $m=12,18,24$ and $30$
represented with circles, squares,
diamonds and triangles respectively. The values of $S$
are normalized by the theoretical values $S_{theo}$ given by the Bose-Einstein ansatz
(\ref{eq5}) for the corresponding initial state $m$.
 }
 \label{fig14}
\end{figure}

Finally, in Fig.~\ref{fig14} 
we present data for the variation of entropy $S$
with the nonlinear parameter $\beta$ 
obtained for different initial states $m$.
The data are averaged over a certain time interval.
The ratio of numerical $S$ values
to the expected theoretical values $S_{theo}$,
given by the Bose-Einstein ansatz (\ref{eq5}),
has a sharp growth at a certain $\beta = \beta_c \approx 1.5$
indicating a thermalization transition at this $\beta_c$.
At the same time there is a certain spreading of curves
which we attribute to fluctuations and the fact that longer time intervals
are required for larger $m$ values where nonlinear 
frequencies become smaller (see discussion of this point
in Sections above). Thus due to a restricted time of our numerical
simulations we cannot exclude that instead of 
a sharp transition to thermalization
there is a crossover  in a certain $\beta$ interval.
In any case the data of Fig.~\ref{fig14} show that for $\beta >3$ 
the values of entropy $S$ become close to the expected theoretical values
with $S/S_{theo} =1$. Thus we conclude that the dynamical thermalization
is reached for $\beta > 2 - 3$.

\section{Discussion}

The results of this work demonstrate the emergence of dynamical 
thermalization of BEC described by the GPE equation
of wave function time evolution in  Sinai oscillator  trap.
The classical dynamics of rays in such a trap is chaotic 
and the quantum properties of this system in absence of interactions
are described by well known results of the field of quantum chaos.
The dynamical thermalization appears above a certain critical strength
of interaction $\beta > \beta_c$. In this thermalized phase the
probability distribution $\rho_m$ over linear eigenmodes (at $\beta=0$)
is well described by the standard statistical Bose-Einstein distribution.
We stress that the dynamical thermalization appears
in a completely isolated system without any external noise.
We point that this thermal distribution is drastically
different from energy equipartition over modes
which is usually expected for nonlinear oscillator 
lattices, including the Fermi-Pasta-Ulam problem, leading to the
ultra-violet catastrophe. Thus our results show that the energy 
is redistributed only over certain low energy modes
and that there is no energy flow to high energy modes.
We think that this result may have interesting implications
to the dynamical consideration of Kolmogorov
turbulence which assumes the presence of energy flow
from large (low energy) to small (high energy)
spacial scales in presence of noise \cite{zakharov,nazarenko}.
Our results indicate that in absence of noise such 
energy flow can be absent due to absence of energy equipartition. 
Of course, further investigations of this system
in numerical simulations on larger time scales 
are highly desirable in view of numerical difficulties
discussed in the previous Section.

The trap configuration considered here had been already
realized experimentally in 3D \cite{ketterle1995}
and we believe that further experimental
investigations of dynamical thermalization 
in the Sinai oscillator trap are accesible 
for modern experiments with cold atoms and BEC.
The variation of interaction strength between atoms 
by means of Feshbach resonance 
should be able to detect a transition from
quasi-integrable phase at $\beta < \beta_c$ to
the phase of dynamical thermalization at 
$\beta > \beta_c$.  The case of Litium 6 atoms,
where the interactions can be changed in a broad range
(see e.g. \cite{salomon}), can be a good test bed for
the studies of dynamical thermalization in the Sinai oscillator trap
and fundamental origins of thermalization in isolated systems.

We thank Pavel Chapovsky and David Gu\'ery-Odelin for 
useful discussions of cold atom physics.


\end{document}